\documentclass[twocolumn]{aastex61}

\usepackage{amsmath}
\usepackage{float}

\accepted{October 19, 2017}
\submitjournal{ApJ}

\begin{document}

\title{Fast and Luminous Transients from the Explosions of Long-lived Massive White Dwarf Merger Remnants}

\author{Jared Brooks}
\affiliation{Department of Physics, University of California, Santa Barbara, CA 93106, USA}

\author{Josiah Schwab}
\affiliation{Department of Astronomy and Astrophysics, University of California, Santa Cruz, CA 95064, USA}
\affiliation{Hubble Fellow}

\author{Lars Bildsten}
\affiliation{Department of Physics, University of California, Santa Barbara, CA 93106, USA}
\affiliation{Kavli Institute for Theoretical Physics, Santa Barbara, CA 93106, USA}

\author{Eliot Quataert}
\affiliation{Astronomy and Physics Departments and Theoretical Astrophysics Center, University of California, Berkeley, CA 94720, USA}

\author{Bill Paxton}
\affiliation{Kavli Institute for Theoretical Physics, Santa Barbara, CA 93106, USA}

\author{Sergei Blinnikov}
\affiliation{M.V. Lomonosov Moscow State University, Sternberg Astronomical Institute, Universitetsky~pr.,~13,~Moscow, 119234, Russia}
\affiliation{Institute for Theoretical and Experimental Physics (Kurchatov Inst., ITEP), 117218 Moscow, Russia}
\affiliation{Kavli Institute for the Physics and Mathematics of the Universe (WPI), The University of Tokyo, Kashiwa, 277-8583,  Japan}

\author{Elena Sorokina}
\affiliation{M.V. Lomonosov Moscow State University, Sternberg Astronomical Institute, Universitetsky~pr.,~13,~Moscow, 119234, Russia}
\affiliation{Institute for Theoretical and Experimental Physics (Kurchatov Inst., ITEP), 117218 Moscow, Russia}

\begin{abstract}

We study the evolution and final outcome of long-lived (${\approx}10^5$ years) remnants from the merger of a He white dwarf (WD) with a more massive C/O or O/Ne WD.  
Using Modules for Experiments in Stellar Astrophysics ($\texttt{MESA}$), we show that these remnants have a red giant configuration supported by steady helium burning, adding mass to the WD core until it reaches $M_{\rm core}\approx 1.12-1.20 M_\odot$.  
At that point, the base of the surface convection zone extends into the burning layer, mixing the helium burning products (primarily carbon and magnesium) throughout the convective envelope.  
Further evolution depletes the convective envelope of helium, and dramatically slows the mass increase of the underlying WD core. 
The WD core mass growth re-initiates after helium depletion, as then an uncoupled carbon burning shell is ignited and proceeds to burn the fuel from the remaining metal-rich extended envelope.  
For large enough initial total merger masses, O/Ne WD cores would experience electron-capture triggered collapse to neutron stars (NSs) after growing to near Chandrasekhar mass ($M_{\rm Ch}$).  
Massive C/O WD cores could suffer the same fate after a carbon-burning flame converts them to O/Ne.  
The NS formation would release ${\approx}10^{50}$ ergs into the remaining extended low mass envelope. 
Using the STELLA radiative transfer code, we predict the resulting optical light curves from these exploded envelopes. 
Reaching absolute magnitudes of $M_V\approx -17$, these transients are bright for about one week, and have many features of the class of luminous, rapidly evolving transients studied by Drout and collaborators.

\end{abstract}


\section{Introduction}\label{sec:intro}

The merger of two WDs is thought to have a wide range of possible outcomes depending on the mass and composition of the WDs \citep[e.g.,][]{Webbink1984, Iben1984, Dan2014, Shen2015}.  
In this work, we study the outcome of the merger of a He WD with a massive C/O or O/Ne WD.  
While such binary systems necessarily have lower mass ratios $(M_{\rm donor}/M_{\rm accretor} \la 2/3)$, they may still merge on contact due to weak spin-orbit coupling during the direct impact accretion phase \citep{Marsh2004, Brown2016} or due to dynamical friction within the expanding ejected shell from a H nova \citep{Shen2015}.  
The mergers of He WDs with C/O WDs of mass $\approx 0.6 M_{\odot}$ are believed to form the hydrogen-deficient, carbon-rich supergiant R Coronae Borealis (RCB) stars \citep{Clayton2012, Clayton2013}; in this scenario, the merger leads to a stably burning helium shell on top of the more massive WD core and this configuration endures for the timescale over which the burning shell consumes the massive He envelope (${\approx}3\times10^5$ yr).

In the case of higher mass C/O WDs, many have investigated the possibility of a helium detonation occurring during the merger and leading to a subsequent detonation of the C/O core \citep{Guillochon2010, Woosley2011, Pakmor2013, Shen2014, Dan2015}.
We are considering an alternative evolution where the large mass of helium forms a giant envelope surrounding the massive (C/O or O/Ne) WD core that burns for ${\approx}10^5$ years.  
Our calculations show the possibility of an unusual explosive outcome upon reaching a near Chandradeskhar mass ($M_{\rm Ch}$) core.

Models of stars with helium-burning shells on top of cold, degenerate (WD-like) cores have been previously constructed.  
Evolutionary calculations have focused on modeling the lower mass RCB stars and so have not considered massive O/Ne cores \citep{Weiss1987, Saio2002, Zhang2012, Menon2013, Zhang2014}.  
Static models have been used to explore helium-shell burning configurations at higher core masses \citep{Biermann1971, Jeffery1988, Saio1988b}, but models with significant envelopes and core masses $\ga 1.1-1.2 M_\odot$ were reported to be difficult to construct.  
This paper is the first to study the evolution of the merger remnants of double WDs that have total masses close to or greater than the Chandrasekhar mass ($M_{\rm Ch}$) and whose degenerate cores can grow to $M_{\rm Ch}$ via stable helium shell burning and later C shell burning.  
We focus on O/Ne ($1.10-1.20 M_\odot$) WDs merging with $0.40 M_\odot$ He WDs.  
We additionally evolve merger remnants of He WDs with massive C/O ($0.86-1.0 M_\odot$) WDs as a means to further study the He shell burning process on massive WD cores. 

In models with O/Ne WD cores that reach core masses near $M_{\rm Ch}$, electron captures in the center will lead to a collapse of the core to a neutron star (NS) in a process very similar to accretion induced collapse (AIC) and electron-capture supernovae (ECSNe) \citep{Nomoto1979, Miyaji1980, Nomoto1987, Canal1990, Nomoto1991, Woosley1992, Ritossa1996, Dessart2006, Metzger2009, Darbha2010, Piro2013, Takahashi2013, Tauris2013}.
We show here that the expected explosion energies of ${\approx}10^{50}$ ergs \citep{Kitaura2006, Dessart2006} and envelope masses ${\approx}0.1 M_\odot$ imply that the resulting transients should be luminous ($L>10^{43}$ erg s$^{-1}$) and rapidly evolving, similar to the class of rapidly evolving transients identified by \cite{Drout2014}.
Other possible members of this class are 2002bj \citep{Poznanski2010} and 2010X \citep{Kasliwal2010}.
We use the recent $\texttt{MESA}$ integration \citep{Paxton2017} of STELLA to generate light curves from our more massive models and compare them to the objects in \cite{Drout2014}.

In \S \ref{sec:grow}, we discuss the growth of massive WD cores and show that the He burning layer eventually couples to the convection zone.
We explore the post-coupling evolution and uncoupled C shell burning in \S \ref{sec:post}.
Then in \S \ref{sec:shock} we use STELLA to generate light curves from our more massive models and compare them to the objects in \cite{Drout2014}.
We discuss the future of simulations and observations of these types of objects in \S \ref{sec:concl}.

\section{Growth of the Degenerate Core}\label{sec:grow}

Using the stellar evolution code \texttt{MESA} (r9793) \citep{Paxton2011, Paxton2013, Paxton2015, Paxton2017}, we constructed idealized models of WD merger remnants by creating C/O WDs of masses $0.86, 0.92,$ and $1.00 M_\odot$ and O/Ne WDs of masses $1.10$ and $1.20 M_\odot$ through the same methods as in \cite{Brooks2016, Brooks2017}.
We add a $0.40 M_\odot$ envelope composed of 98\% $^4$He, 1\% $^{14}$N, and 1\% other metals, corresponding to the approximate core compositon of a solar metallicity He WD of this mass.
We then relax the envelope to a constant entropy of $10^9$ erg cm$^{-3}$ K$^{-1}$, and the core to a constant temperature equal to the center temperature after 10 Myr (for consistency between models) of cooling after WD formation, typically $T_c=3-7\times10^7$ K.
This envelope entropy and core temperature prescription approximates a configuration immediately following a merger \citep{Benz1990, Dan2014, Schwab2016}.
We do not include any rotation in our models.
Viscous stresses efficiently transport angular momentum within the merger remnant and quickly lead the remnant to a quasi-spherical, thermally-supported state \citep{Shen2012, Schwab2012}.  
As the envelope expands in response to the thermal energy deposited by the merger, a small amount of mass shed from large radii can remove most of the angular momentum, leaving the envelope slowly rotating; this motivates our choice of non-rotating models.  
Our simple, ad hoc initial conditions will be erased after few thermal timescales and since our focus is on the nuclear timescale evolution of these objects, they provide a suitable starting state.

\begin{figure}[H]
  \centering
  \includegraphics[width = \columnwidth]{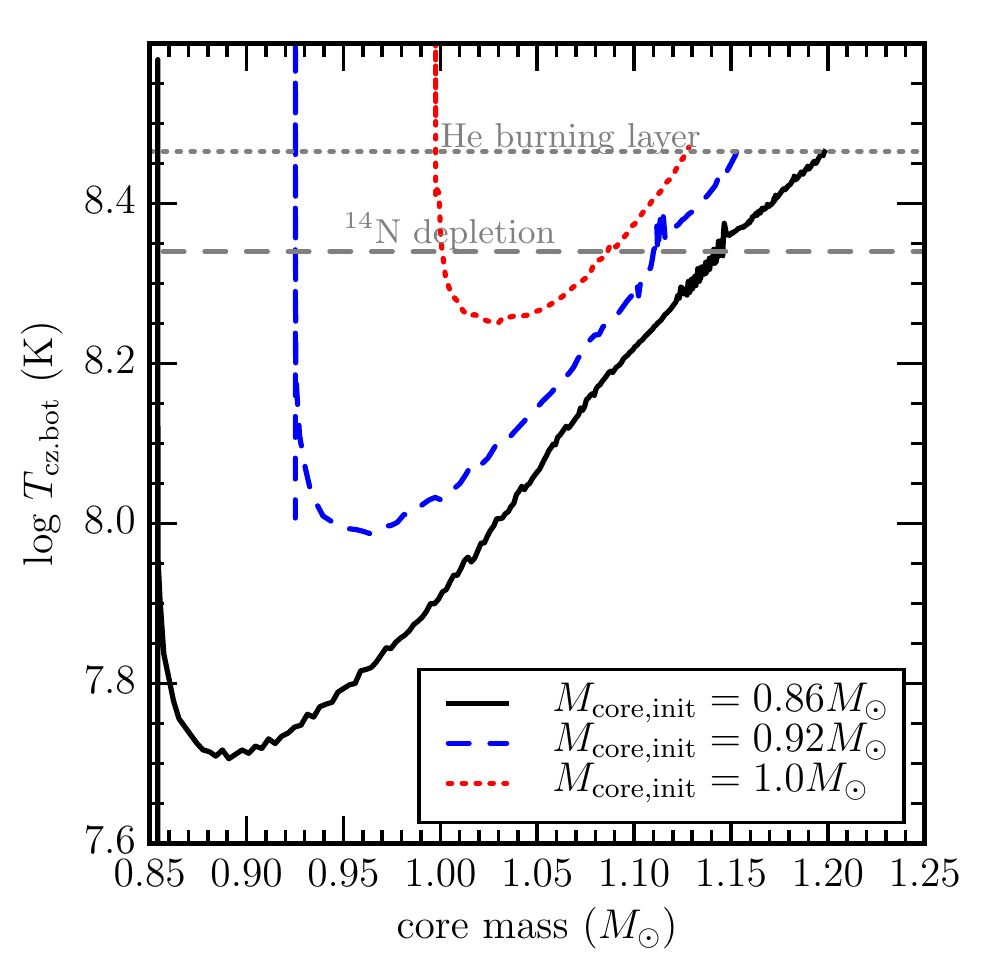}
  \caption{\footnotesize Temperature at the base of the convection zone as a function of core mass for three models with initial core masses of $0.86, 0.92$, and $1.00 M_\odot$.
  The temperature at which $^{14}$N is depleted from the envelope is shown by the grey dashed line.
  The temperature of the helium burning layer is shown by the grey dotted line.
  The burning layer and convective zone effectively couple when the $T_{\rm cz.bot}=T_{\rm He\, burn}$.}
  \label{fig:2}
\end{figure}

We then evolve the models forward in time.
Due to the initially compact configuration, the models start with a nuclear burning flash at the base of the core/envelope interface that expands the envelope into a red giant.
After this initial flash, the models steadily burn He and deposit the C/O ashes onto the core. 
Much like a H shell burning red giant branch star, this helium burning layer powers an overlying convection zone that extends out to the surface.
Our calculations make use of low-temperature opacities ($\log T/{\rm K} < 4.5$) that include the effects of carbon enhancement.  
We generate a set of tables with X = 0, Z = 0.02, and a range carbon enhancement factors (varying from 1 to 100) using the {\AE}SOPUS web interface \citep{Marigo2009}.
We incorporate these tables in \texttt{MESA} using its \texttt{other\_kap} hook.

\subsection{\texttt{MESA} Models}

Since the He burning shell is being fed by the large He envelope, instead of accretion from a binary companion, the He shell burns at the maximum rate allowed by the core mass-luminosity relation \citep{Kippenhahn1981, Jeffery1988}, which coincides with the maximum steady helium burning rate \citep[${\approx} 3-4\times10^{-6} M_\odot$/yr]{Piersanti2014, Brooks2016}.
For the model that starts with a core mass of $0.86 M_\odot$, the burning has settled from the initial flash and becomes steady when the core has grown to $0.90 M_\odot$ (black solid line in Figure \ref{fig:2}).
At that point, the model has a surface luminosity of $L_{\rm surf}=37,600 L_\odot$ compared to a burning luminosity of $L_{\rm burn}=31,300 L_\odot$, with an effective temperature of $T_{\rm eff}=13,500$ K, radius of $35 R_\odot$, and surface opacity of $\kappa_{\rm surf}=0.16$ cm$^2$/g, which agree well with the models from \cite{Saio1988, Jeffery1988}.
As the core mass grows, the convection zone extends deeper in to higher temperatures.
In $10^5$ years this model reaches a core mass of $M_{\rm core}=1.20 M_\odot$, with $T_{\rm eff}=12,800$ K and a radius of $63 R_\odot$.

In Figure \ref{fig:2} we show the temperature at which $^{14}$N is depleted from the envelope.
We derive this temperature by integrating the rate of alpha captures onto $^{14}$N over the entire convective envelope and finding the temperature at which the timescale of $^{14}$N depletion from the envelope is $10^4$ yr, approximately 5\% of the lifetimes of these stars.
The $^{14}$N depletion timescale is calculated via
\begin{equation} \dfrac{1}{t_{14}} \equiv \dfrac{d\ln X_{14}}{dt} = \dfrac{Y}{4 m_p M_{\rm conv}}\int_{M_{\rm conv}}^{M_*} \rho\langle\sigma v \rangle dM .\end{equation}
Performing the same calculation for $^{18}$O shows that its depletion temperature is only 5\% higher than that for $^{14}$N, so there is negligible time to see an enhancement of $^{18}$O through this mechanism.

During this phase of helium shell burning, the models studied in this paper share similarities to R Coronae Borealis (RCB) stars \citep{Clayton2012}.  
Several RCB stars have been observed to have significantly enhanced amounts of $^{18}$O on their surfaces, which favors the double WD merger scenario \citep{Clayton2007, Menon2013}.
The number of known RCB stars then implies a WD merger rate of 1 per 100 years in our galaxy, consistent with population synthesis expectations and observed double degenerate binaries \citep{Alcock2001, Jeffery2011, Zhang2012, Zhang2014, Karakas2015, Brown2016}.  
By measuring the semi-regular stellar pulsations from RCB stars, their masses are derived to be in the range of ${\sim}0.8-0.9 M_\odot$ \citep{Saio2008}, which agrees well with the binary population synthesis from \citet{Han1998}.

Due to their higher core masses, the models shown here are at least a factor of a few more luminous than the typical RCB star.  
The higher $T_{\rm eff}$ of the models means they would likely be better observationally classified as extreme He stars \citep{Jeffery2008a, Jeffery2008}.  
As indicated in Fig.~\ref{fig:2} and the related discussion, the higher temperatures at the base of the convective envelope would lead them to appear depleted in $^{14}$N and $^{18}$O for core masses $\ga 1.05 M_{\odot}$.

The hot C/O ashes of He burning settle onto the core at such a high rate that a shell ignition of carbon occurs in the freshly accreted C/O ash layer at core masses of $1.2-1.3 M_\odot$ \citep{Brooks2016}, depending on the starting core mass and temperature.
The ignited carbon develops into a flame that propagates inwards, which we do not follow. 
But this presumably converts the entire C/O core into O/Ne \citep{Nomoto1985, Lecoanet2016}.
For an initially O/Ne core, this flame converts the C/O shell to O/Ne \citep{Brooks2017}.

Hence, the cores of all massive WD merger remnants will be devoid of central carbon when the core mass grows to $M_{\rm Ch}$.
For systems that start with a C/O WD primary, only the most massive systems have total masses $\gtrsim M_{\rm Ch}$, with the most massive case studied here having a total mass of $1.40 M_\odot$.
The carbon flames in models with C/O cores will lift the core degeneracy upon reaching the center.
Merger remnants with masses $\lesssim M_{\rm Ch}$ may complete nuclear burning at this stage and end up as cooling O/Ne WDs.
Although these models, and those of lower mass, do not lead to the types of explosions we are interested in for this paper, they are useful in helping us describe the peculiar behavior of He burning shells on WD cores of masses $M_{\rm core}\approx 1.12-1.20 M_\odot$.
Merger remnants with initially C/O cores and total masses $\gtrsim M_{\rm Ch}$ will grow cores massive enough that during Kelvin Helmholtz (KH) contraction they will ignite Ne off-center and follow the evolution described in \cite{Schwab2016}. 
This involves the formation of a low mass Fe core, but still ends with the collapse to a NS.

\subsection{Core Masses $> 1.1 M_\odot$}

We find that as the degenerate cores of our models grow in mass, the base of the convection zone extends deeper into the model (see Figure \ref{fig:2}) until it overlaps with the He burning shell when $M_{\rm core}\approx 1.12-1.20 M_\odot$ (the exact core mass at which this coupling happens depends on the history of the star, e.g. its initial core mass and temperature).
Once the burning layer couples with the convective envelope, the burning products are no longer primarily deposited on the core, but a large portion of them are mixed with the entire envelope.

To understand why the convection zone base extends deeper into the star until it reaches the burning layer, we first must explain why the envelope is convective in the first place.
We see in Figure \ref{fig:2} that even at core masses of $0.90 M_\odot$ the base of the convection zone extends to $\log T/{\rm K}=7.7$.  
The reason for this is that He burning occurs at much higher temperatures and densities and lower opacities than H burning; high enough temperature that the Klein-Nishina relation becomes relevant \citep{Buchler1976}.
This means that opacity is increasing outwards from the burning layer as 
\begin{equation}\label{eq:1} \kappa_{\rm es} \approx 0.2\left[1+\left(\dfrac{T}{4.5\times10^8 \text{ K}}\right)^{0.86}\right]^{-1} \text{cm}^2\text{g}^{-1}  .\end{equation}

\begin{figure}[H]
  \centering
  \includegraphics[width = \columnwidth]{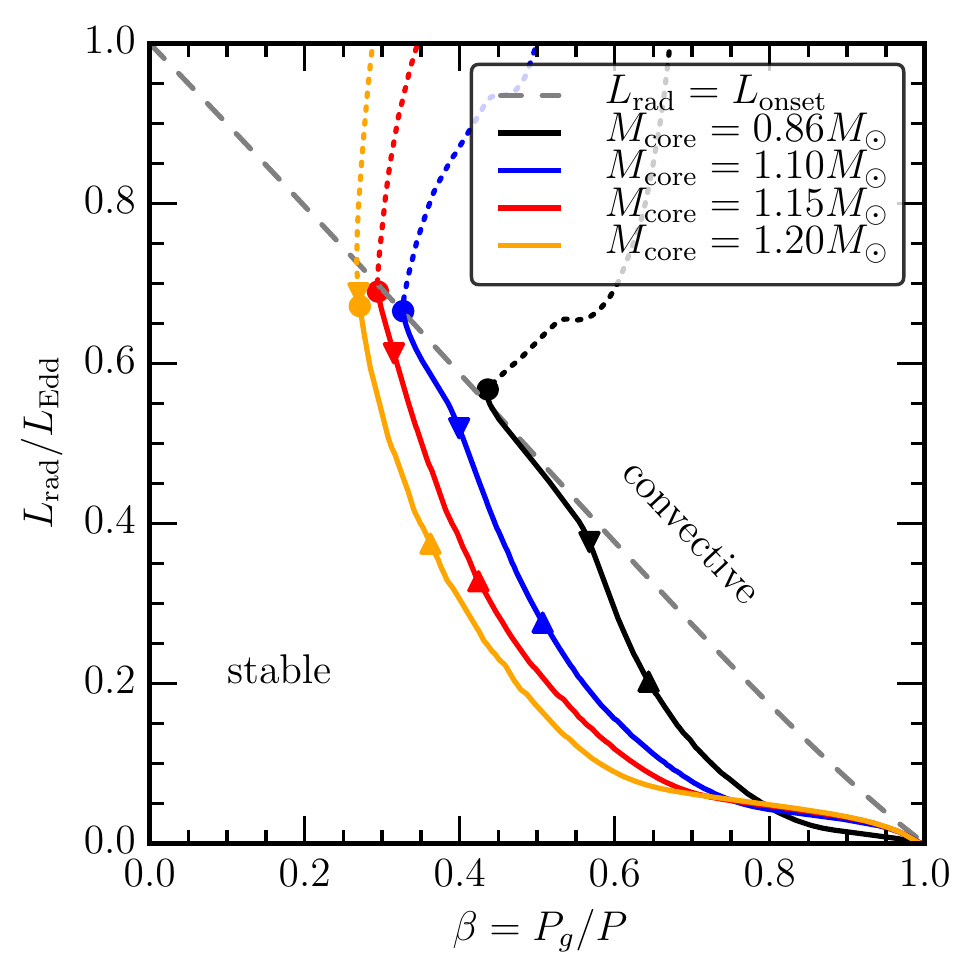}
  \caption{\footnotesize Profiles from the model with an initial core mass of $0.86 M_\odot$ are shown at various later core masses.
  The grey dashed curve is the convective instability line described by equation \ref{eq:99}.
  The ``up'' triangles are where $L(r)=0.5 L_{\rm surf}$, the ``down'' triangles are where $L(r)=0.9 L_{\rm surf}$.
  The circles mark the base of the convection zone, and the profiles are solid in radiative regions and dotted in convective regions.
  Note how the ``down'' triangle on the last profile sits above the base of the convective layer, implying that burning is occurring  in the convective envelope.}
  \label{fig:4}
\end{figure}

\noindent Furthermore, the core mass-luminosity relation is steeper for He shell burning sources than for H shell burning sources \citep{Jeffery1988}.
This means that for a given core mass, at the burning layer, the He shell source will have higher radiation luminosity ($L_{\rm rad}$) and a smaller Eddington luminosity ($L_{\rm Edd}$) that decreases at larger radius.
As we move outwards from the burning layer and $\kappa$ is increasing, we encounter a point where $L_{\rm rad}/L_{\rm Edd}$ exceeds the convective instability criterion described in \cite{Joss1973, Paxton2013}:
\begin{equation}\label{eq:99} \dfrac{L_{\rm rad}}{L_{\rm Edd}} > \dfrac{8(1-\beta)(4-3\beta)}{32-24\beta+3\beta^2} , \end{equation}
where $\beta=P_g/P$, shown by the grey dashed line in Figure \ref{fig:4}.
This shows that the reason for deep convection in these models is that the local luminosity just above the burning layer is approaching $L_{\rm Edd}$.

Since we know that the burning luminosity increases with $M_{\rm core}$ more rapidly than $L_{\rm Edd}$ does \citep{Kippenhahn1981, Jeffery1988}, the ratio $L/L_{\rm Edd}$ must increase with $M_{\rm core}$.
This means that as the core mass grows and we follow the $L_{\rm rad}=L_{\rm onset}$ grey dashed line up to higher values of $L_{\rm rad}/L_{\rm Edd}$, the base of the convection zone must have a decreasing $\beta$ with increasing core mass.
The convection zone base moves inwards to higher temperatures and thus lower values of $\beta$.

\begin{figure}[H]
  \centering
  \includegraphics[width = \columnwidth]{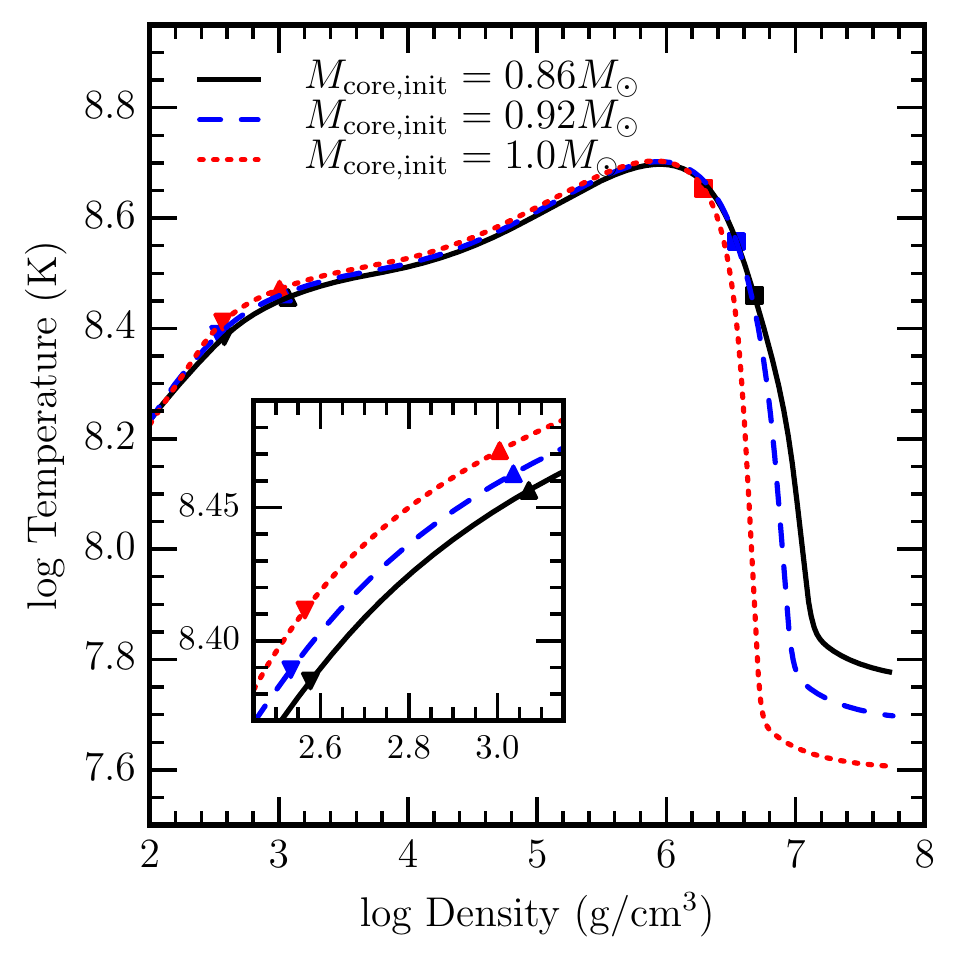}
  \caption{\footnotesize Profiles from different initial core masses at a common core mass of $1.10 M_\odot$.
  The ``up'' triangles are where $L(r)=0.5 L_{\rm surf}$, the ``down'' triangles are where $L(r)=0.9 L_{\rm surf}$.
  The squares mark the initial core mass coordinates.
  This plot emphasizes the effect of initial conditions on the density and temperature of the burning layer at a common core mass.}
  \label{fig:1}
\end{figure}

As the convection zone base moves deeper into higher temperatures, the mass between the burning layer and convection zone becomes smaller, which steepens the profiles in Figure \ref{fig:4}, as there is a smaller region where both $L(r)$ and $\kappa(r)$ are relatively constant.
This means that as $M_{\rm core}$ grows, the value of $L_{\rm rad}/L_{\rm Edd}$ at the base of the convection zone increases, which decreases the value of $\beta$ at the base of the convection zone, which decreases the mass between the burning layer and convection zone (where both $L(r)$ and $\kappa(r)$ are relatively constant). 
This, in turn, steepens the profiles in Figure \ref{fig:4} leading to smaller values of $\beta$ in the burning layer.
As $\beta$ decreases (radiation pressure increases) in the burning layer, the entropy rises rapidly until it is comparable with the entropy of the convection zone.
At this point, the convective envelope couples to the burning layer, mixing a significant fraction of the burning products into the envelope.

As we mentioned above and show in Figure \ref{fig:2}, models with different starting masses experience the coupling between the burning layer and the convective envelope at different core masses.
This is because this coupling is sensitive to the temperature, density, and the exent of the burning layer, which in turn depend on the history of the model.
For a given core mass, models with higher initial core masses have less time for conduction to heat the core and have less freshly-burned ashes directly below the burning layer.
This leads to different core temperatures, luminosities, and radii for models with different initial core masses when compared at a common core mass (Figure \ref{fig:1}).

From this point on in the paper, we will only discuss models that start with O/Ne WD cores, such that they will begin runaway electron captures in their centers upon reaching $M_{\rm Ch}$.
Models that instead start with C/O WD cores will ignite carbon shell flames before reaching $M_{\rm Ch}$ that convert the entire C/O core to O/Ne and lift the core degeneracy, following the evolution described in \cite{Schwab2016}.
Even though such an evolution still may lead to the collapse of a degenerate core to a NS inside an extended envelope, the envelope masses will be much smaller (due to small initial C/O core masses).
The resulting lightcurves would only last ${\lesssim}$a couple days, and are thus harder detect.
We defer exploration of these explosions and light curves to future work.

\section{Post-coupling Regimes}\label{sec:post}

As discussed above, when the core mass grows large enough to cause the burning layer to couple with the convective envelope, the burning products are evenly mixed throughout the convective envelope instead of primarily being deposited on the core.
This then causes uniform depletion of helium in the envelope, and a rising metallicity.
At the thermodynamic conditions present in the hot parts of the envelope, the default configuration of the \texttt{MESA} EOS module uses the OPAL equation of state for $Z \le 0.04$ \citep{Rogers2002}.  
The OPAL tables do not extend to higher $Z$, and so for more metal-rich mixtures the \texttt{MESA} EOS switches to using the Helmholtz EOS \citep[henceforth HELM, ][]{Timmes2000}, which does not self-consistently include ionization.  
For the models presented in this paper we use the OPAL tables for $Z < 0.08$, using the $Z=0.04$ table for $0.04 < Z < 0.08$.  
We use the HELM EOS, assuming complete ionization, for $Z > 0.10$.  
We blend between the two over the interval $0.08 < Z < 0.10$.  
With this choice, we observe that the convective zone and burning layer remain coupled until the helium is largely depleted from the envelope.
Remaining on the OPAL tables at higher Z continues to include the effects of He-ionization in the He-dominated envelope, at the cost of underestimating the effect of the ionization of metals.  
Delaying the switch to HELM until higher Z allowed us to separate the onset of the fully-coupled evolution from this EOS change, and to ensure that the EOS blend was not adversely affecting the model\footnote{\texttt{MESA} provides an option to schematically include ionization when using HELM by blending from a version of HELM assuming complete ionization to one assuming a fully neutral composition.
The location and extent of this blend (in temperature) is user-specified.
We experimented with several blend locations, but all models that used this option were plagued by convergence errors caused by the blend.
We were unable to evolve any of these models for sufficiently long durations to allow their core masses to grow to $M_{\rm Ch}$.}.

Due to the large core masses (and small core radii), the helium burns at such high temperatures and densities that the $^{16}$O nuclei produced are quickly consumed by $\alpha$-captures, along with the $^{20}$Ne nuclei, such that the composition of the envelope becomes $^{24}$Mg rich, with significant amount of $^{12}$C (mass fractions given in Table \ref{tab:1}).

\begin{figure}[H]
  \centering
  \includegraphics[width = \columnwidth]{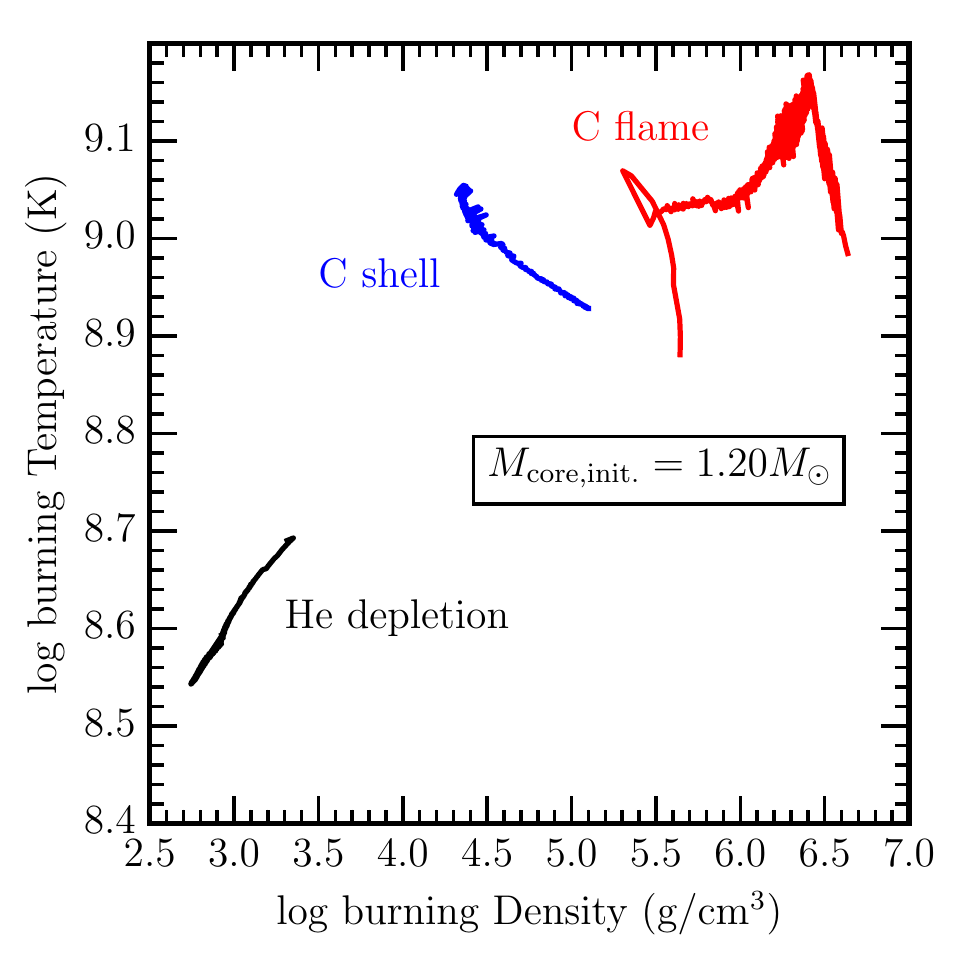}
  \caption{\footnotesize Evolution of the main burning layer during different stages for the model with $M_{\rm core, init.}=1.20 M_\odot$.
  Due to the initially large core mass, the first phase of evolution depletes the envelope of He (black curve), followed by a C-flame (red curve) leading to C-shell burning (blue curve) powering the outer convection zone while keeping the radius $>400 R_\odot$.}
  \label{fig:3}
\end{figure}

\begin{figure}[H]
  \centering
  \includegraphics[width = \columnwidth]{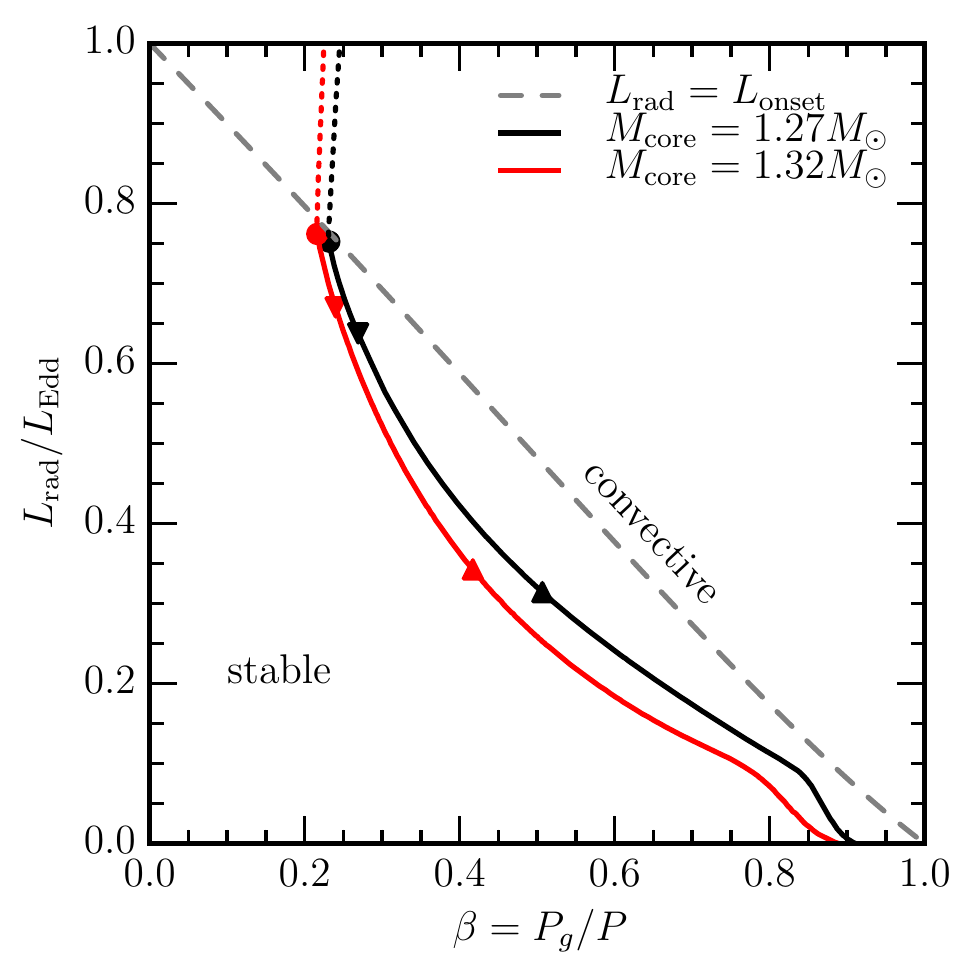}
  \caption{\footnotesize Same as Figure \ref{fig:4} but for a coupled simulation that burns all of its He and starts doing C-shell burning.
  These two profiles are from the C-shell burning phase.
  Note that during C shell burning, the burning layer and convective envelope are not coupled, so that C burning grows the mass of the degenerate core, leading to AIC.}
  \label{fig:6}
\end{figure}

Once there is insufficient helium to power a burning layer to prop up the envelope, it begins to Kelvin-Helmholtz (KH) contract until the underlying carbon is ignited, and a carbon flame propagates through the existing C/O layer until it reaches the O/Ne core \citep{Brooks2017}, as shown by the red curve in Figure \ref{fig:3}.
Once the flame is extinguished, the envelope continues to KH contract until the Mg/C layer reaches densities and temperatures high enough for carbon burning, shown by the blue curve in Figure \ref{fig:3}.
This carbon shell burning powers the convective envelope and keeps the radius above $400 R_\odot$.
The carbon shell burning occurs at temperatures ($T=8-11\times10^8$ K) and densities ($\rho=2-13\times10^4$ g/cm$^3$) high enough that, even though the temperature at the base of the convection zone was just as hot as it was during helium depletion and only gets hotter as the core grows, the carbon burning layer remains decoupled from the convection zone (see Figure \ref{fig:6}).
This means that the composition in the envelope remains constant and uniform, as the underlying O/Ne core grows in mass, eventually triggering an AIC.
Hence, for each super-Chandrasekhar mass model, there is a Mg/C envelope of mass ${\approx}0.1 M_\odot$ above the O/Ne core, with a density structure shown in Figure \ref{fig:7}.

\section{Shock Heating the Envelopes}\label{sec:shock}   

The collapse of the core to a NS releases a large amount of gravitational potential energy, which can be transformed to thermal/kinetic energy to power a bright transient.
The steep density gradient present in our models is similar to that around O/Ne and low mass Fe cores produced in single star evolution which are known to robustly lead to collapse to a NS and generate a shock energy of ${\approx}10^{50}$ ergs \citep{Kitaura2006, Melson2015, Radice2017}.
The light curve will look different from that of canonical AIC \citep{Woosley1992, Piro2013} or a stripped-envelope electron capture supernova \citep{Moriya2016}, since the envelopes in our models extend out to $\sim$ a few$\times100 R_\odot$, and thus produce much brighter light curves.
They also look different from regular electron-capture SNe from $8-10 M_\odot$ stars, due to the much smaller envelope masses, producing shorter light curves that lack hydrogen \citep{Takahashi2013, Smith2013, Tauris2015, Jones2016}.

\begin{figure}[H]
  \centering
  \includegraphics[width = \columnwidth]{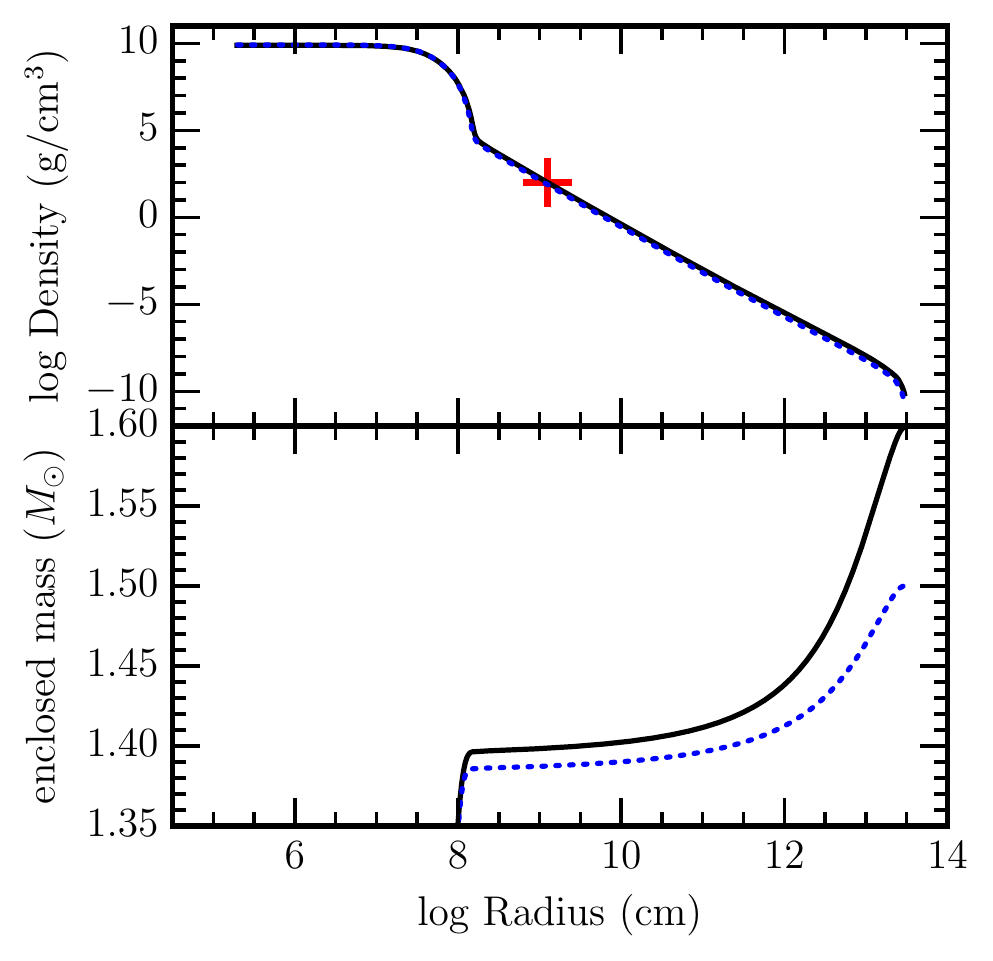}
  \caption{\footnotesize Profiles from the two models used for our explosion calculations when the core reaches conditions for AIC.
  Top Panel: Density structure.
  The red cross marks the mass coordinate at which we separate the core zones from the envelope zones and inject the SN energy.
  Bottom Panel: Mass profile, showing that most of the mass of the envelope is above the energy injection coordinate.}
  \label{fig:7}
\end{figure}

We take the most massive models described in \S \ref{sec:grow} and \ref{sec:post} after reaching core masses near $M_{\rm Ch}$ such that electron captures in the core began \citep{Schwab2015}, and we remove the $1.4 M_\odot$ cores just above the burning layer where $\log T/K=8.3$, $\log\rho/{\rm g\,cm}^{-3} =2$, and $r=0.018 R_\odot$ (marked by the red cross in Figure \ref{fig:7}), and inject $10^{50}$ ergs of energy into the inner $0.01 M_\odot$ of the remaining envelope over $5$ milliseconds.

The resulting shock then propagates through the envelope for about a third of a day, and when it is near the surface, we save data from the model and use it as input for the \texttt{STELLA} code.
\texttt{STELLA} is an implicitly differenced hydrodynamics code that incorporates multigroup radiative transfer \citep{Blinnikov1993, Blinnikov2006}.
\texttt{STELLA} typically uses about 100 frequency groups, enough to produce a spectral energy distribution, but not sufficient to produce spectra.
The opacity is computed based on over 153,000 spectral lines from \citet{Kurucz1995} and \citet{Verner1996}.  Opacity also includes photoionization, free-free absorption and electron scattering.

We use the two most massive envelope models, a $0.1 (0.2) M_\odot$ envelope from the $1.5 (1.6) M_\odot$ total mass model that started with a $1.1 (1.2) M_\odot$ O/Ne WD core.
At the time of explosion, the envelopes extended out to $410 R_\odot$ and $426 R_\odot$ for the $0.1 M_\odot$ and $0.2 M_\odot$ envelopes, respectively.
The envelopes' composition is uniform and dominated by $^{24}$Mg and $^{12}$C, with mass fractions given in Table \ref{tab:1}.

\begin{deluxetable}{ccccccc}
  \tablewidth{0.5\textwidth}
  \tablecaption{Elemental Mass Fractions in the envelopes at explosion\label{tab:1}}
  \tablehead{\colhead{$M_{\rm env}/M_\odot$} & \colhead{$^{24}$Mg} & \colhead{$^{12}$C} & \colhead{$^{20}$Ne} & \colhead{$^{16}$O}  & \colhead{$^{28}$Si} & \colhead{$^4$He}}
  \startdata
  0.1  & 0.603   & 0.308   & 0.032   & 0.017  & 0.015  &0.000\\
  0.2  & 0.653   & 0.269   & 0.022   & 0.017  & 0.014  &0.000\\
  \enddata
  \tablecomments{\footnotesize For both of the models studied in this section, we give\\ the mass fractions of the dominant elements. Composition is\\ uniform throughout the envelope for both models.}
\end{deluxetable}

\begin{figure}[H]
  \centering
  \includegraphics[width = \columnwidth]{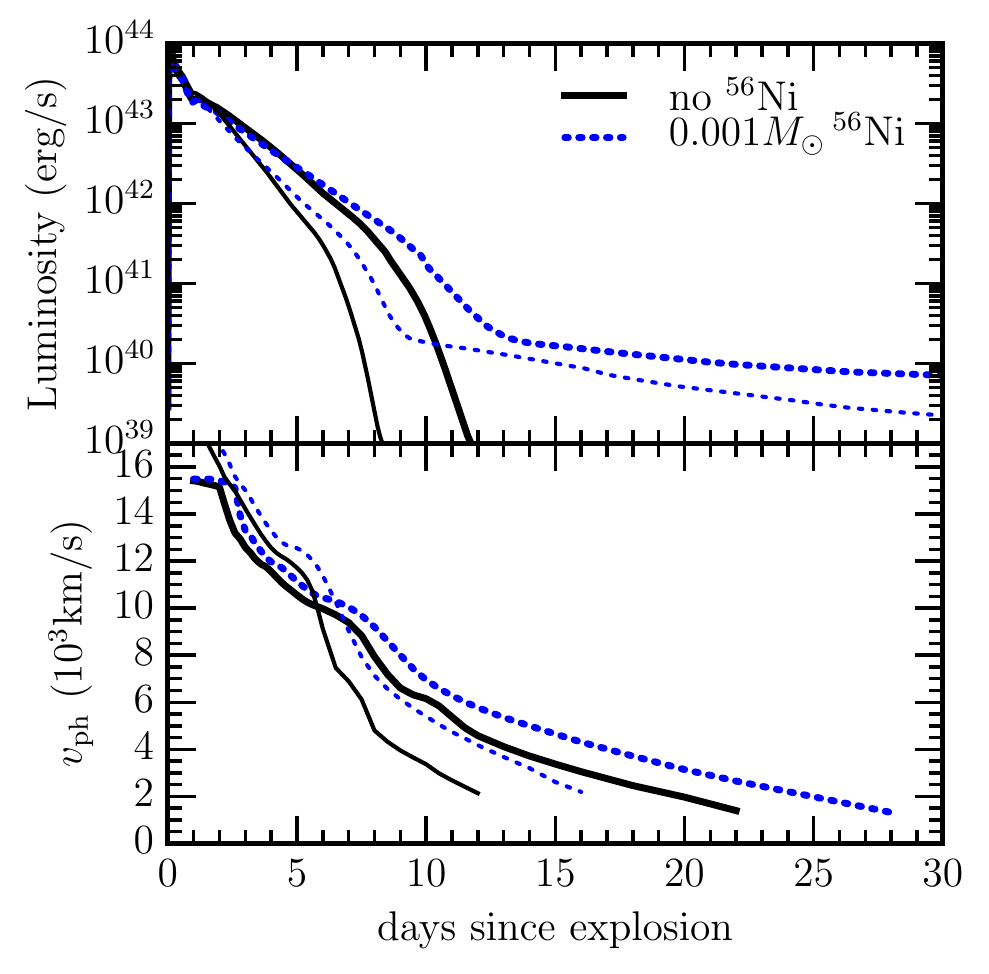}
  \caption{\footnotesize Bold curves are for the $0.2 M_\odot$ model. Thinner curves are for the $0.1 M_\odot$ model.
  Top panel: bolometric luminosity.
  Bottom panel: Photospheric velocites.}
  \label{fig:11}
\end{figure}

The light curves for these two models are shown in the top panel of Figure \ref{fig:11}, with the photospheric velocities shown in the bottom panel.
Figure \ref{fig:12} shows additional properties of these models.
The reason for the short light curves and high photospheric velocities is the low ejecta mass; we are injecting SN energies into an envelope with a small fraction of the mass of the regular shock-powered type IIP SNe.
The peak luminosity is that expected for the radiative losses of an expanding and cooling ejecta, while the duration is consistent with the time it takes for the ejecta to become optically thin.

The light curves through the filters B, V, I, and R show (Figure \ref{fig:13}) that the peak in optical light occurs 3-4 days post-explosion for the $0.2 M_\odot$ model.
The rise time of the optical light curves is slightly faster than the decay time, but a modest amount of $^{56}$Ni may slow the decay time to be more consistent with the objects in \cite{Drout2014}.
The $^{56}$Ni would be generated by the intense nuclear burning behind the shock near the core-envelope boundary just after core collapse.
We artificially add the $^{56}$Ni behind the shock front in our models since we do not trust our 1D calculation to accurately predict the right amount of $^{56}$Ni.
An addition of $0.001 M_\odot$ of $^{56}$Ni in the inner $0.03 M_\odot$ extends the visible light curve by ${\approx}$a day (shown in Figure {\ref{fig:13}).
We found that $0.001 M_\odot$ of $^{56}$Ni was the minimum amount necessary to keep the bolometric luminosity above $10^{40}$ erg/s at day 20 after energy injection.
Compare this to the ${\approx}2.5\times10^{-4} M_\odot$ of $^{56}$Ni found in AIC calculations by \cite{Dessart2006}, but with only a total ejecta mass of $0.001 M_\odot$, because these models did not have the extended envelope considered here.
Additionally, we also varied the injection energy.
A factor of 2 increase in the explosion energy leads to half of a magnitude increase in peak brightness and ${\approx}$a day decrease in light curve duration, and vice versa for a factor of 2 decrease in explosion energy.

\begin{figure}[H]
  \centering
  \includegraphics[width = \columnwidth]{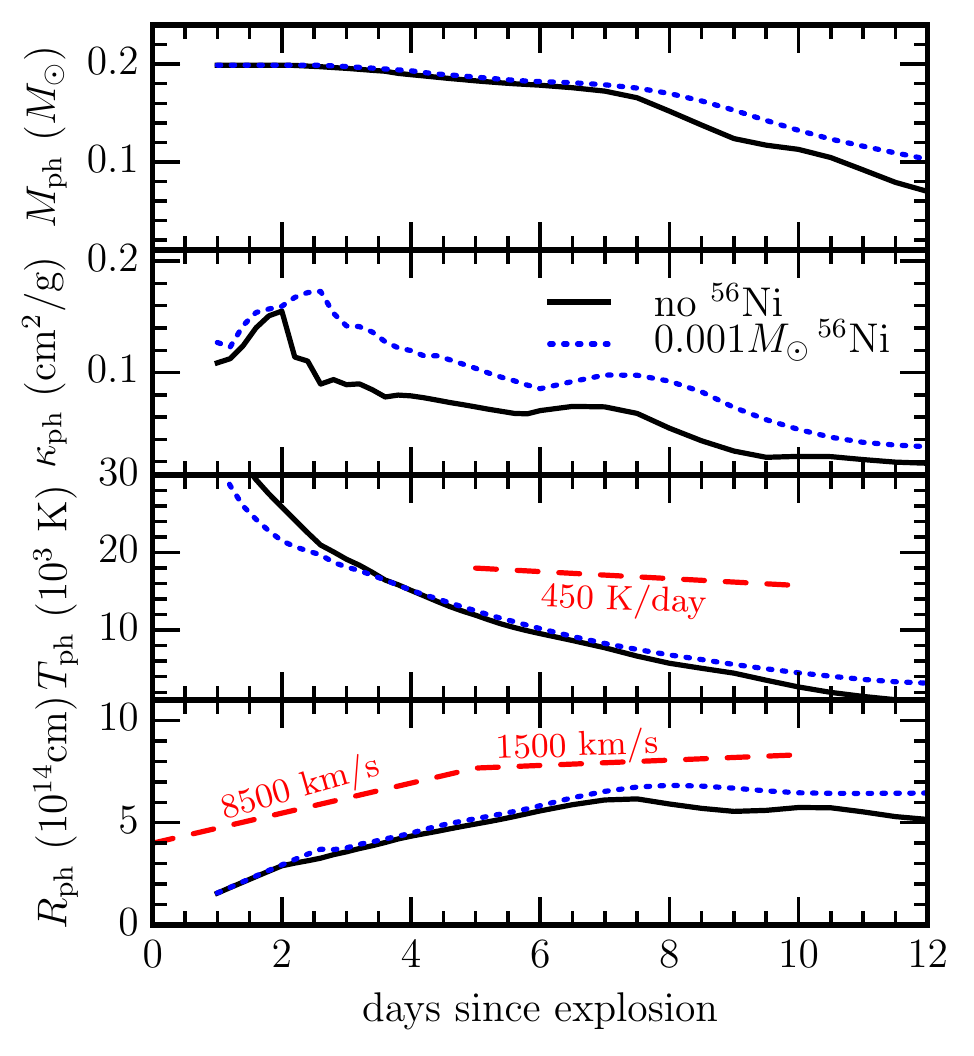}
  \caption{\footnotesize Additional properties from our STELLA radiation transfer calculations of explosions of the $0.2 M_\odot$ envelope with no $^{56}$Ni (solid black curve) and with $0.001 M_\odot$ of $^{56}$Ni (dotted blue curve), same as in Figure \ref{fig:11}.
  Top panel: mass coordinate of the photosphere.
  Second panel: Opacity at the photosphere.
  Third panel: Temperature of the photosphere.
  Bottom panel: Radius coordinate of the photosphere.
  The red dashed lines show fiducial cooling rates and photosphere expansion speeds based of the best observed object from \cite{Drout2014}: PS1-10bjp.}
  \label{fig:12}
\end{figure}

From the selection criteria given in \S 2.2 of \cite{Drout2014}, both of the explosion models shown here would be included in this new class of rapidly evolving luminous transients \citep[see also ][for additional observational examples]{Arcavi2016, Tanaka2016}.
Specifically, the transients rise by $\gtrsim 1.5$ mag in the 9 days immediately prior to observed maximum light, and they decline by $\gtrsim 1.5$ mag in the ${\sim}25$ days post observed maximum.
The third selection criterion is that the transient must be present in at least three sequential observations.
The relatively short light curves of these models, compared to the average timescales of transients in \cite{Drout2014}, somewhat reduces the likelihood of obtaining three or more sequential observations.

\begin{figure}[H]
  \centering
  \includegraphics[width = \columnwidth]{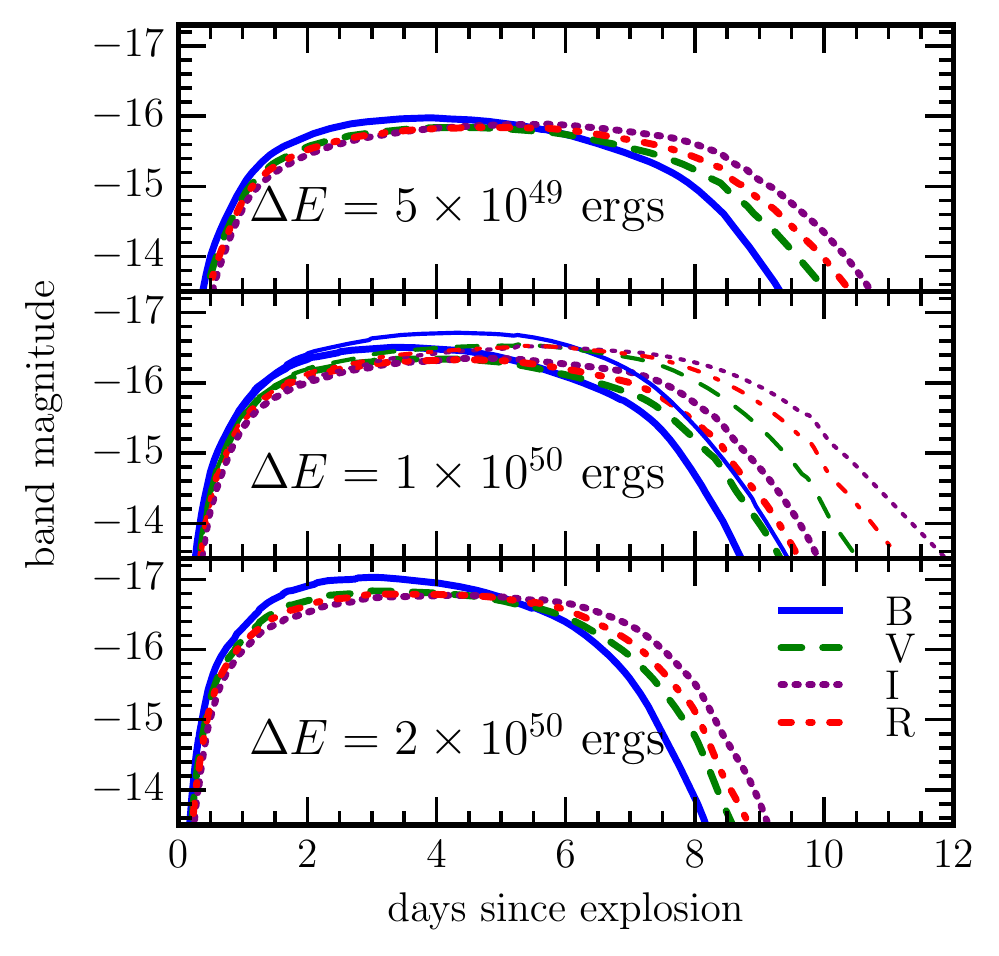}
  \caption{\footnotesize B, V, I, and R band light curves for Mg/C envelopes of $0.2 M_\odot$ with explosion energies of $5\times10^{49}$ ergs(top panel), $1\times10^{50}$ ergs (middle panel), and $2\times10^{50}$ ergs (bottom panel).
  The middle panel also includes a simulation with $1\times10^{50}$ ergs and $0.001 M_\odot$ of $^{56}$Ni in the inner $0.03 M_\odot$ of the envelope (thin lines).
  }
  \label{fig:13}
\end{figure}

\section{Conclusions}\label{sec:concl}

We have presented the first evolutionary simulations of super-Chandrasekhar merger remnants of He and O/Ne WDs.
The remnants expand to giant configurations supported by He burning shells that power extended convection zones.
We have shown that upon reaching core masses $M_{\rm core}\approx 1.12-1.20 M_\odot$, depending on initial conditions, the base of the surface convection zone extends deep enough into the burning layer that the burning products are no longer primarily deposited on the core, but mixed evenly throughout the convective envelope.
This leads to He depletion and metal enrichment of the envelopes, resulting in uniform compositions dominated by $^{24}$Mg and $^{12}$C (see Table \ref{tab:1}).
Once the He is depleted and the Mg/C envelopes begin to KH contract, they ignite and steadily burn the $^{12}$C in a shell, uncoupled to the convection zone, keeping the envelope extended above $400 R_\odot$ when the core reaches conditions for electron capture induced collapse.

Our $1.6 M_\odot$ model spends ${\approx}7000$ years in the C shell burning phase prior to reaching AIC conditions, with a luminosity of $7\times10^{38}$ erg/s.
This is quite bright, so finding a pre-SN progenitor associated with the resulting transient would be feasible within about 10 Mpc.
The merger rate producing such systems is uncertain but the super-Chandrasekhar merger rate in the Milky Way is perhaps one every 1000 years \citep{Badenes2012} so that there could be several such C shell burning objects in the Galaxy and M31 at any time.    
Observationally, these would appear as very luminous H deficient C stars, with an unusually high abundance of $^{24}$Mg.   
The low $T_{\rm eff}$ mean that dust production might well be efficient, possibly leading to large mass loss rates in dust-driven winds and significant self-obscuration \citep{Schwab2016}.   
One significant uncertainty in our evolutionary calculations is that dust-driven winds can in principle be strong enough to decrease the mass of the remnant below $M_{\rm Ch}$, avoiding the collapse and explosion scenario we have explored in this paper.   
Unfortunately, the physics of dust production and dust-driven winds under these conditions are not well enough understood to robustly determine whether or not the remnants will remain above $M_{\rm Ch}$.

If the cores reach $M_{\rm Ch}$, electron captures in the center will lead to the collapse of the degenerate core to a NS.
The resulting shock propagates through the envelope and generates a light curve as the ejecta expand.
The light curve (Figures \ref{fig:11} and \ref{fig:13}) is bright ($>10^{43}$ erg s$^{-1}$) due to the extended ($>400 R_\odot$) envelope and short (${\sim}$a week) due to the low ejecta mass (${\approx}0.1 M_\odot$).

These fast and luminous light curves fit many of the observed features of a class of rapidly evolving luminous transients \citep{Drout2014}, including peak brightness ($L>10^{43}$ erg s$^{-1}$), light curve shape and duration (time above half-maximum of $<12$ days), radius and velocity evolution, and hot continuum-dominated spectra.
There are other objects that have been suggested as progenitors to the rapidly evolving and luminous transients, including stripped massive stars \citep{Tauris2013a, Kleiser2014}, ``.Ia'' SNe from detonations of helium shells on WDs \citep{Bildsten2007, Shen2010, Perets2010}, and shock breakout from a dense circumstellar shell \citep{Ofek2010}.
Due to the very fast light curves and velocities generated by the exploding WD merger remnant models shown here (Figures \ref{fig:11}-\ref{fig:13}), we conclude that they can only account for a subset of this observed class of rapidly evolving, luminous transients.
The lack of H and He and strong enrichment of $^{12}$C and $^{24}$Mg in the spectra for these objects may help to distinguish them from other possible fast and luminous transients.

In our calculations, we chose to study systems with $0.4 M_\odot$ He WDs, as this is roughly the maximum mass of He WDs. 
This is useful for us because we are interested in systems with a total mass above $M_{\rm Ch}$. 
In addition, higher ejecta masses correlate with longer and brighter light curves that more closely match the new transients described in \cite{Drout2014}.
The effect of varying the accreted He mass is comparable to varying the initial core mass: the evolutionary path will be slightly different, as shown in Figures \ref{fig:2} and \ref{fig:1}, but the resulting structure and composition of the envelopes will be very similar, as shown in Figure \ref{fig:7} and Table \ref{tab:1}.
Future work should further explore the parameter space of He + O/Ne WD binaries, drawing on guidance from observations and population synthesis calculations, in order to more fully characterize these intriguing outcomes.

We thank Maria Drout for useful discussions. 
We acknowledge stimulating workshops at Sky House where these ideas germinated. 
This work was supported by the National Science Foundation through grant PHY 11-25915 and ACI 16-63688 and  funded in part by the Gordon and Betty Moore Foundation through Grant GBMF5076.
Support for this work was also provided by NASA through Hubble Fellowship grant \# HST-HF2-51382.001-A awarded by the Space Telescope Science Institute, which is operated by the Association of Universities for Research in Astronomy, Inc., for NASA, under contract NAS5-26555. 
EQ is supported in part by a Simons Investigator award from the Simons Foundation and the David and Lucile Packard Foundation.
Many of the simulations for this work were made possible by the Triton Resource, a high-performance research computing system operated by the San Diego Supercomputer Center at UC San Diego.

\software{MESA \citep{Paxton2011, Paxton2013, Paxton2015, Paxton2017}, {\AE}SOPUS \citep{Marigo2009}, OPAL \citep{Rogers2002}, HELM \citep{Timmes2000}, STELLA \citep{Blinnikov1993, Blinnikov2006}}

\bibliographystyle{apj}
\bibliography{wdmerger}

\end{document}